\documentclass[lettersize,journal]{IEEEtran}
\usepackage{amsmath,amsfonts,amssymb,amsthm,mathrsfs}
\usepackage{algorithmic}
\usepackage{algorithm}
\usepackage{array}
\usepackage[caption=false,font=normalsize,labelfont=sf,textfont=sf]{subfig}
\usepackage{textcomp}
\usepackage{stfloats}
\usepackage{url}
\usepackage{verbatim}
\usepackage{graphicx}
\usepackage{cite}
\usepackage{threeparttable}
\usepackage{booktabs}
\usepackage{nicematrix}
\usepackage{multirow}
\usepackage{xcolor}
\usepackage{enumitem}

\begin{document}

% [STATUS: working title. The current title is aligned with the new narrative,
% but may still be shortened for WCL once the numerical section is finalized.]
\title{Covering-radius and Collinearity- Minimizing Pilots  for Channel Estimation in TDD Systems
%Geometry-Aware Joint Time-Frequency Pilot Pattern Design for Channel Estimation in TDD Systems
}

\author{Xu Zhu, Yi Zeng, and Tiejun Li\thanks{X. Zhu, Yi Zeng, and T. Li are with Laboratory of Mathematics and Applied Mathematics, School of Mathematical Sciences,  Center for Machine Learning Research, Peking University, Beijing 100871, P.R. China}
\thanks{Email: xuzhu@pku.edu.cn (X. Zhu),  zengyi0427@pku.edu.cn (Y. Zeng), tieli@pku.edu.cn (T. Li)}
\thanks{Corresponding author: Tiejun Li}}

\markboth{IEEE Wireless Communications Letters,~Draft}{}

\IEEEpubid{}

\maketitle

\begin{abstract}
This letter studies pilot design for orthogonal frequency-division multiplexing-based time-division duplex (TDD) systems under a sliding-window latest-slot recovery framework that jointly exploits delay--Doppler sparsity across recent slots. Under contiguous-subband and fairness constraints, this viewpoint naturally leads to a geometry-aware time--frequency joint pilot assignment. We show that effective patterns should balance grid coverage and redundant-collinearity suppression, with an additional symmetry-avoidance refinement when complete collinearity elimination is infeasible. Based on these principles, we formulate a mixed-integer construction method compatible with practical TDD allocation. Numerical results show that minimum-coverage-radius and collinearity-control (MCC) pattern improves both surrogate geometry metrics and latest-slot recovery performance.
\end{abstract}


\section{Introduction}\label{sec:intro}

% \IEEEPARstart{S}{parse} recovery or compressed sensing have been widely applied to channel estimation, and the associated pilot-design problem has also been extensively studied in the literature \cite{P1_bajwa2010compressed,P2_berger2010application}. In OFDM systems, channel recovery is typically based primarily on the accuracy of single-slot compressed sensing. Accordingly, pilot patterns are mostly designed from an in-slot viewpoint, with the goal of supporting reliable recovery while maintaining fairness across users \cite{P3_pakrooh2012ofdm,P5_qi2014pilot,WYB_wan2025multi}. This single-slot viewpoint underlies much of the existing pilot-design methodology.

\IEEEPARstart{S}{parse} recovery or compressed sensing have been widely applied to channel estimation, and the associated pilot-design problem has also been extensively studied in the literature \cite{P1_bajwa2010compressed,P2_berger2010application}. In orthogonal frequency-division multiplexing (OFDM) systems, when each user is allocated a pilot set that is not overly limited, accurate channel recovery can often be achieved from single-slot observations alone, provided that the in-slot pilot positions are properly designed at the subcarrier level \cite{P3_pakrooh2012ofdm,P5_qi2014pilot,WYB_wan2025multi}. As a result, much of the existing pilot-design methodology is developed from a slot-wise viewpoint, emphasizing in-slot sensing quality and user fairness.

In practical time-division duplex (TDD) systems, however, pilot allocation is often constrained jointly by protocol and physical-layer requirements. Pilots are typically transmitted in contiguous frequency-domain blocks, and when many users are simultaneously served, each user may only occupy a relatively narrow pilot subband. As a result, recovering the full-band channel from the current slot alone becomes a large-bandwidth frequency-extrapolation problem. In slowly varying scenarios, it may therefore be preferable to carry historical channel information forward across time, while using the current-slot pilots mainly to refine the occupied subband.

\begin{figure}[t]
\centering
\includegraphics[width=\columnwidth]{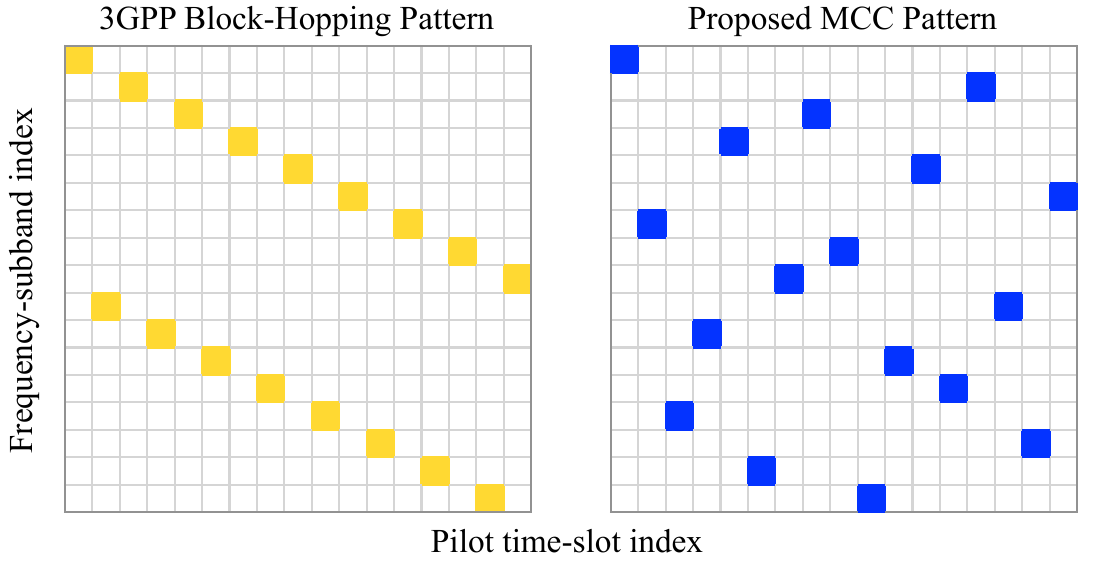}
\caption{Illustration of pilot base patterns on the $k\times k$ time--frequency grid ($k=17$). Each colored square indicates an active subband--time pair $(f,t)$ with $X_{ft}=1$. Left: conventional 3GPP SRS-based block-hopping pattern, whose pilot blocks follow a regular linear trajectory across slots. Right: proposed MCC pattern, which spreads the pilot blocks more evenly and avoids overly repetitive collinear structures.}
\label{fig:pilot_patterns}
\end{figure}

A natural question is then whether better reconstruction performance can be achieved by jointly exploiting the two-dimensional sparsity in the delay--Doppler (DD) domain. This observation has appeared in related delay--Doppler frameworks, especially orthogonal time frequency space (OTFS), where joint DD-domain processing is more explicit \cite{P7_OTFS,P8_raviteja2019embedded}. However, the gain brought by DD-domain joint sparse recovery does not itself require a new modulation format. In this letter, we keep the underlying OFDM system unchanged and adopt a sliding-window latest-slot recovery viewpoint. Specifically, for a window of length $T$, the pilot observations from the most recent $T$ slots are jointly used to reconstruct the DD-domain sparse channel, which is then mapped to the full-band channel of the latest slot.

Once the recovery target is changed in this way, pilot design must also be reconsidered. The relevant object is no longer an isolated in-slot pilot arrangement, but a time--frequency pattern over the whole observation window. Consequently, a regular pilot-block hopping rule that is convenient under conventional protocol design, including the 3GPP SRS-based frequency-hopping configuration specified in TS 38.211 \cite{3gpp_ts38211_v1850} (Clause 6.4.1.4.3), can become unfavorable when judged by the multi-slot sensing criterion induced by joint recovery. Guided by the channel-extrapolation geometry and mutual-coherence considerations, we propose the MCC pilot, emphasizing \textit{minimum coverage} radius together with \textit{collinearity} control. The resulting design principles admit a tractable mixed-integer linear programming formulation under contiguous-subband and fairness constraints for constructing pilot patterns.

\section{System Model and Design Objective}\label{sec:system}

\subsection{TDD Pilot Allocation Pattern}\label{subsec:constraints}

We consider a simplified time--frequency resource model. At each time instant, the base station uses $N$ subcarriers and assigns a contiguous subband of length $M$ to each active user. Let $k = N/M$ be an integer, so that at most $k$ users can be simultaneously supported. Define the subbands $\mathcal B_f := \{(f-1)M+1,\ldots,fM\}, \quad (f=1,\ldots,k).$ At each time slot, the collection $\{\mathcal B_f\}_{f=1}^k$ is assigned to the $k$ users without overlap.

To encode the pilot arrangement, we introduce a binary matrix $X\in\{0,1\}^{k\times k}$, where the column index corresponds to time and the row index corresponds to the subband index. The feasibility conditions are
\begin{equation}
    \sum_{f=1}^k X_{ft}=1,\ \forall t;
    \qquad
    \sum_{t=1}^k X_{ft}=1,\ \forall f.
    \label{eq:rc_norm_new}
\end{equation}
Thus, each time slot activates exactly one subband in the base pattern, and each subband is visited exactly once over one period. For user fairness, we adopt a protocol-inspired cyclic strategy: a single base pattern is designed, and different users use cyclic column shifts of that pattern. This ensures that all users share the same geometry up to a deterministic shift. Fig.~\ref{fig:pilot_patterns} illustrates two representative base patterns on the $k\times k$ grid (with $k=17$ in Fig. \ref{fig:pilot_patterns}); in the figure, each colored square represents an active location $(f,t)$ with $X_{ft}=1$.

% \keepnote{This subsection is a direct revision of your original ``System Model and Pilot Allocation Constraints''. The main content is preserved, but the exposition is tightened for WCL.}

\subsection{Windowed Joint-Recovery Model}\label{subsec:joint_model}

Pilot design in this letter is driven by the following recovery task. For a latest slot $t_0$ and a short observation window $\Omega_{t_0}:=\{t_0-T+1,\ldots,t_0\},$ we jointly use the pilot observations within $\Omega_{t_0}$ to reconstruct the latest-slot full-band channel of user $u$. A compact observation model is
\begin{equation}
    Y_t^{(u)} = P_t^{(u)}\Phi_t^{(u)}H_t^{(u)} + W_t^{(u)},
    \qquad t\in \Omega_{t_0},
    \label{eq:obs_model}
\end{equation}
where $P_t^{(u)}$ encodes the subband selected by the pilot pattern, $\Phi_t^{(u)}$ denotes the pilot waveform or modulation operator, $H_t^{(u)}$ is the channel response at slot $t$, and $W_t^{(u)}$ is noise.

When the channel evolution over $\Omega_{t_0}$ is moderate, one may use a DD-domain sparse representation of the form $H_t^{(u)} = F\tilde H^{(u)}G_t^{\top},$ where $F$ is the frequency--delay partial Fourier matrix, $G$ is the time--Doppler partial Fourier matrix, and $G_t$ denotes the row associated with slot $t$. This leads to the standard sparse regularized fitting problem
\begin{equation}
    \tilde H_\text{est}^{(u)}
    =
    \arg\min_{\tilde H}
    \sum_{t\in \Omega_{t_0}}
    \left\|
    Y_t^{(u)}-P_t^{(u)}\Phi_t^{(u)}F\tilde HG_t^{\top}
    \right\|_2^2
    +\lambda\|\tilde H\|_1,
    \label{eq:joint_lasso}
\end{equation}
and the newest-slot full-band channel is then recovered as $H_{\text{est},t_0}^{(u)}=F\tilde H_\text{est}^{(u)}G_{t_0}^{\top}$. In practice, the window size $T$ is usually modest, so \eqref{eq:joint_lasso} remains computationally realistic by the essential joint sparse structure. Eq.~\eqref{eq:joint_lasso} serves as the representative joint-recovery task used to evaluate different pilot patterns and can be enhanced by leveraging more sophisticated priors \cite{Yufei2026DDA}.

% We emphasize this model here because it is the concrete surrogate through which different pilot patterns are evaluated in our study; moreover, once \eqref{eq:joint_lasso} is established, more sophisticated OMP-type, statistical, or algorithm-unrolling solvers can all be viewed as operating on the same windowed recovery task.

\subsection{Why Multi-Slot Recovery Changes Pilot Design}\label{subsec:why_multislot}

Given the windowed recovery task in \eqref{eq:joint_lasso}, pilot design differs from the conventional slot-wise setting in two main aspects.

First, the pilot pattern now lives on a two-dimensional time--frequency grid, so purely slot-wise optimality is no longer sufficient. A commonly used 3GPP-style block-wise hopping rule is illustrated in the left panel of Fig.~\ref{fig:pilot_patterns} for odd $k$. To see why such a structure can become problematic under joint recovery, consider the virtual-domain correlation induced by the full-window pilot pattern. Let $f_t$ denote the active pilot-block index at time $t$. For a uniformly hopped pattern as in the left panel of Fig.~\ref{fig:pilot_patterns}, these block locations satisfy $f_{t+1}-f_t \equiv d \pmod{k}$ for some fixed hop increment $d$. Although this rule is regular and implementation-friendly in each slot, its joint sensing geometry over multiple slots can be highly unfavorable. In particular, the normalized DD-domain correlation kernel satisfies
\begin{equation}
\label{eq:legacy_kernel}
|K(\tau,\nu)|
=\frac{1}{Mk}
\bigg|
\frac{\sin\!\big(k\pi(\nu-dM\tau)\big)}
     {\sin\!\big(\pi(\nu-dM\tau)\big)}
\bigg|\cdot
\bigg|
\frac{\sin(M\pi\tau)}
     {\sin(\pi\tau)}
\bigg|,
\end{equation}
where the normalization is by the main peak value at $(\tau,\nu)=(0,0)$. Hence large correlation values concentrate near the ridge $\nu-dM\tau\equiv 0 \pmod 1$. Moreover, the strongest off-origin ambiguity occurs at $(\tau,\nu)\equiv \pm(1/(Mk),\, d/k)\pmod 1$, where \eqref{eq:legacy_kernel} evaluates to $\sin(\pi/k)\big/(M\sin(\pi/(Mk)))$, which approaches $1$ as $M$ and $k$ grow. Therefore, a legacy pattern that is benign from a per-slot viewpoint may become highly coherent under joint multi-slot recovery. As a contrast, chirp-type pilot \cite{P9_Chirp} trajectories such as $f_t\equiv t^2 \pmod{k}$  use a deliberately non-regular quadratic arrangement across slots, which reshapes the induced interference pattern and can reduce the maximum grid-point coherence close to the Welch benchmark \cite{P10_welch}. This comparison shows that, once multiple slots are fused, pilot design is no longer a purely slot-wise hopping problem, but a genuinely two-dimensional sensing-geometry problem.

Second, practical recovery does not always use the full observation window, but may instead adopt short temporal sub-windows of different sizes. When the channel evolves slowly, a shorter window is often sufficient and reduces computational cost. On the other hand, an excessively long window may violate the local channel-consistency assumption underlying joint recovery. Therefore, in practical channel tracking, the receiver may adapt the recovery window according to operating conditions, and pilot design should not focus solely on global full-window diversity, but should also avoid locally unfavorable configurations that may severely degrade recovery under certain window choices. Motivated by this, rather than directly optimizing all possible dynamic recovery objectives, we seek a feasible matrix $X$ satisfying \eqref{eq:rc_norm_new} whose time--frequency geometry is favorable for reliable latest-slot full-band recovery and remains robust under varying sub-window sizes. To this end, we introduce two local geometric rules for time--frequency pilot allocation.

\section{Geometry-Guided Design Principles}\label{sec:design}

The pilot pattern is guided by two main principles: 1) the pilots should cover the time--frequency grid as evenly as possible; 2) the pattern should suppress excessive repetition of certain collinear configurations.

\subsection{Coverage-Aware Principle}\label{subsec:coverage}

We first introduce a coverage metric for the pilot pattern. For any feasible pattern $X$ and any metric matrix $C\in\mathbb R_+^{k^2\times k^2}$ on the time--frequency grid, define the distance from grid point $(f,t)$ to its nearest pilot by
\begin{equation}
    a_{ft}(X;C)
    :=
    \min_{f',t'}\big\{ C_{ft\to f't'} : X_{f't'}=1 \big\}.
\end{equation}
A pattern with small $a_{ft}(X;C)$ provides nearby pilot support for the recovery of location $(f,t)$. This is especially relevant when only a shorter sub-window is used, because long extrapolation in frequency or time generally leads to larger sensitivity.

A natural worst-case quantity is the covering radius $r(X;C):=\max_{f,t} a_{ft}(X;C).$ For fixed $(k,C)$, let $r_k$ denote the minimum achievable covering radius over all feasible $X$. We then select pilot patterns by minimizing the average distance while enforcing the optimal worst-case radius:
\begin{equation}
    \min_X \sum_{f,t} a_{ft}(X;C)
    \quad \text{s.t.}\quad
    \eqref{eq:rc_norm_new}\ \&\ a_{ft}(X;C)\le r_k.
    \label{eq:kmedian_stage1}
\end{equation}
This criterion favors patterns that are uniformly spread across the grid without sacrificing the worst-case guarantee.

The same preference for good coverage can also be interpreted from the mutual-coherence viewpoint, in the spirit of \cite{P3_pakrooh2012ofdm,P5_qi2014pilot}. 
For the virtual-domain dictionary without further delay oversampling, the correlation coefficient in the two-dimensional setting can be written as
\begin{equation}
    \rho_{i,j}^2
    = \frac{1}{|\mathcal P|}
    + \frac{2}{|\mathcal P|^2}
    \sum_{(df,dt)\in \mathcal T_{\rm rep}}
    c_{df,dt}\,\Gamma_{df,dt}(i,j),
    \label{eq:coherence_collinear_new}
\end{equation}
where $\Gamma_{df,dt}(i,j):= \cos\!\left(2\pi(df\cdot i+dt\cdot j)/k\right)$, $\mathcal T
:= \{(f'-f,t'-t):(f,t),(f',t')\in\mathcal P,\ (f,t)\neq(f',t')\},$ and $\mathcal T_{\rm rep}\subseteq\mathcal T$ contains one representative from each antipodal pair
$\{(df,dt),(-df,-dt)\}$, while $c_{df,dt}$ denotes the multiplicity of the corresponding difference class. In the exact block-subband model, the same term is further weighted by the delay-direction Dirichlet factor already contained in $K(\tau,\nu)$, which only strengthens the sensitivity to small-delay interactions.

This expression shows that clustered pilots generate many short difference vectors, which tend to concentrate coherence at small virtual-domain offsets $(i,j)$. Here, small $(i,j)$ refers to the relative separation between paths. Since practical channels typically occupy only a limited spread on the virtual delay--Doppler grid, such local coherence concentration reduces the resolvability of nearby paths and makes reconstruction more difficult.

\subsection{Redundant-Collinearity Suppression and Symmetry Refinement}\label{subsec:collinearity}

Coverage alone does not control all harmful sensing interactions. A pilot pattern may be well spread in the coverage sense, yet still exhibit excessive regularity. Such regularity can counterintuitively create large coherence among virtual-domain offsets, particularly in practical TDD settings with $k\leq 20$.

From \eqref{eq:coherence_collinear_new}, the mechanism is intuitive: repeated difference vectors amplify the same oscillatory component and may therefore create large coherence peaks at certain virtual offsets. In the present two-dimensional pilot-allocation problem, this effect is shaped not only by multiplicity but also by geometry. For a fixed virtual offset $(i,j)$, all difference vectors $(df,dt)$ satisfying $df\cdot i + dt\cdot j \equiv 0 \pmod{k}$ contribute with $\Gamma_{df,dt}(i,j)=1$ and thus align maximally. 
This indicates that harmful configurations are associated with redundant sets of difference vectors lying on the same modular line through the origin. Since these vectors are induced by pairs of pilot locations on the time--frequency grid, this naturally motivates controlling pilot collinearity in the observation domain. In particular, by limiting the number of pilots lying on a common modular line, one reduces the occurrence of repeatedly aligned difference vectors.

For practical TDD values of $k$, completely eliminating redundant collinearity, namely configurations with three or more aligned pilots, is often infeasible. Even so, it remains desirable to suppress the most harmful regularity on a modular line. In particular, if several pilot points lie on the same modular line with symmetric spacing, then the induced coherence contributions tend to reinforce one another rather than cancel out. Therefore, when redundant collinearity is unavoidable, we explicitly exclude symmetric pilot triples along the same modular line.

\section{Optimization-Based Pattern Construction}\label{sec:optimization}

The previous section introduced the geometry-guided design principles for robust latest-slot recovery. We now translate them into an implementable surrogate optimization model on the full $k\times k$ time--frequency grid.

\subsection{Surrogate Formulation of the Design Principles}

\paragraph{Coverage Linearization}
In Sec.~\ref{subsec:coverage}, we introduced the nearest-pilot distance $a_{ft}(X;C)$ and the coverage-driven problem \eqref{eq:kmedian_stage1}. A useful feature of \eqref{eq:kmedian_stage1} is that it has the form of a standard $k$-median-type assignment problem and therefore admits the following exact linearization:
\begin{align}
    \min_{X,E}\quad & \sum_{f,t,f',t'} C_{ft\to f't'}E_{ft\to f't'}
    \label{eq:cover_lin_obj},\\
    \text{s.t.}\quad
    & \sum_{f',t'}E_{ft\to f't'}=1,\qquad \forall f,t,\label{eq:cover_lin1}\\
    & E_{ft\to f't'}\le X_{f't'},\qquad \forall f,t,f',t',\label{eq:cover_lin2}\\
    % & X\text{ satisfies }\eqref{eq:rc_norm_new},\label{eq:cover_lin3}\\
    & E_{ft\to f't'}=0\quad \text{if }C_{ft\to f't'}>r_k,\label{eq:cover_lin4}\\
    & X_{ft},E_{ft\to f't'}\in\{0,1\}\ \& \text{ satisfies }\eqref{eq:rc_norm_new}.\label{eq:cover_lin5}
\end{align}
Here, the binary variable $E_{ft\to f't'}$ indicates whether grid point $(f,t)$ is assigned to pilot location $(f',t')$. The restriction \eqref{eq:cover_lin4} removes assignments outside the covering radius $r_k$, so only nearby candidate links need to be retained in the model.

For implementation, we use the asymmetric Manhattan-type metric $C_{ft\to f't'} := |f-f'| + [t-t' \bmod k]$, which favors both frequency proximity and temporally recent pilot support. The asymmetry in time is motivated by the fact that, within a local recovery window, estimating the channel at the newest time index can only exploit pilot information from the current or earlier slots.

\paragraph{Redundant-Collinearity Budget and Symmetry Control}
Section~\ref{subsec:collinearity} shows that harmful coherence is closely related to repeated difference vectors and, more specifically, to redundant collections of differences lying on the same modular line. Directly modeling this mechanism in the difference-vector domain would lead to a much harder quadratic-assignment-type formulation. We therefore adopt an observation-domain surrogate that captures the same geometric effect by limiting the number of modular lines that may contain collinear pilot triples:
\begin{equation}
    \sum_{(f,t)\in l} X_{f,t}\le Z_l+2,\ Z_l\in\{0,1\}\ (\forall l\in\mathcal S);\quad \sum_{l\in\mathcal S} Z_l \le L_r.
    \label{eq:col_penalty}
\end{equation}

When $k$ is prime, at most two pilots are allowed on each modular line $l$ unless the binary variable $Z_l$ is activated, in which case a single collinear triple is admitted on that line. The budget $\sum_{l\in\mathcal S} Z_l \le L_r$ then limits the total number of admissible redundant-collinearity events. We adopt this hard budget rather than a soft regularizer in order to avoid introducing an additional trade-off parameter. When $k$ is composite, the characterization of modular lines by $i\,f + j\,t \equiv c \pmod{k}$ is no longer applicable in general. We therefore restrict attention to modular lines induced by primitive directions $(u,v)$ satisfying $\gcd(u,v,k)=1$, namely $u\, f + v\, t \equiv c \pmod{k}$, where $c=0,\ldots,k-1$. For prime $k$, all aligned vectors $(i,j)$ determine the same modular direction, so there are $k(k+1)$ distinct modular lines. For composite $k$, this number may be slightly larger, depending on the factorization of $k$, so we enumerate the modular lines directly.

For the practical TDD sizes of interest, stronger four-point collinearities can typically still be excluded. When some collinear triples remain unavoidable, we further suppress symmetric multiplicity along the same modular line. For prime $k$, we impose
\begin{equation}
    X_{ft}+X_{f-d,t(l,-d)}+X_{f+d,t(l,+d)}\le 2,\quad (\forall l,d,f,t),
    \label{eq:col_symmetry}
\end{equation}
where $t(l,\pm d)$ denotes the corresponding time index on modular line $l$ when the subband index is shifted from $f$ to $f\pm d$. When $k$ is composite, \eqref{eq:col_symmetry} is interpreted with respect to the primitive-direction-based modular-line representation above.

\subsection{Integrated Mixed-Integer Linear Programming Formulation}\label{subsec:master_problem}

Combining \eqref{eq:cover_lin_obj}--\eqref{eq:col_symmetry}, we obtain the integrated mixed-integer linear programming (MILP)
\begin{equation}
    \min_{X,E,Z}\ \sum_{f,t,f',t'} C_{ft\to f't'}E_{ft\to f't'}
    \quad
    \text{s.t.}\ 
    \eqref{eq:cover_lin1}\text{--}\eqref{eq:col_symmetry}.
    \label{eq:master_model}
\end{equation}
For the TDD regimes of interest, \eqref{eq:master_model} remains a moderate-scale $0$--$1$ mixed-integer linear programming and can therefore be solved directly by Gurobi. In implementation, the budget $L_r$ can be selected by gradually tightening the admissible redundant-collinearity level until further tightening becomes infeasible or noticeably harms coverage.

\section{Numerical Validation}\label{sec:numerics}

We evaluate the sliding-window latest-slot recovery task on 3 km/h CDL-B channels generated by the Matlab 5G Toolbox. Unless otherwise stated, each plotted value is the median over 200 channel realizations. Since a feasible pilot pattern may appear with any of its $k$ cyclic shifts in practice, robustness across shifts is also important. Therefore, for each channel realization, we evaluate all $k$ cyclic shifts of the pattern, compute the latest-slot NMSE for each shift, select the worst $\lfloor k/4\rfloor$ values, and average them. The plotted performance is then the median of this worst-quarter average over the 200 realizations. In this way, the reported curves reflect not only the average recovery accuracy but also robustness to unfavorable cyclic shifts.

For each fixed cyclic shift, \eqref{eq:joint_lasso} is solved by FISTA~\cite{FISTA} with 500 iterations, together with virtual-domain refinement and Doppler-spread-based truncation. The regularization parameter is set as
$\lambda=\sqrt{2\bar\sigma_{t_0}^2 N_\tau N_\nu\log(N_\tau N_\nu)/(MT)}$,
where $\bar\sigma_{t_0}^2:=\frac1T\sum_{t\in\Omega_{t_0}}\sigma_t^2$, $N_\tau$ and $N_\nu$ denote the numbers of delay and Doppler grid points, respectively. We compare the proposed MCC pilot with the following baselines: 1) a regular 3GPP SRS-based block-hopping pattern in Fig.~\ref{fig:pilot_patterns}, given by $f_t\equiv f_0+t\lfloor k/2\rfloor \pmod{k}$ for odd $k$, and by the interleaved progression $f_t\equiv f_0+\lfloor t/2\rfloor +(k/2)(t\bmod 2)\pmod{k}$ for even $k$; 2) the chirp pattern $f_t\equiv t^2 \pmod{k}$ introduced in Sec.~\ref{subsec:why_multislot}; 3) random feasible patterns, generated independently for each channel realization; and 4) two ablations of the proposed design, namely ``coverage only'' and ``collinearity only.'' All optimized pilot patterns are solved to zero optimality gap, except that the MCC pilot for $k=19$ is terminated at a dual gap of $0.05$.

\begin{figure}[t]
\centering
\includegraphics[width=\columnwidth]{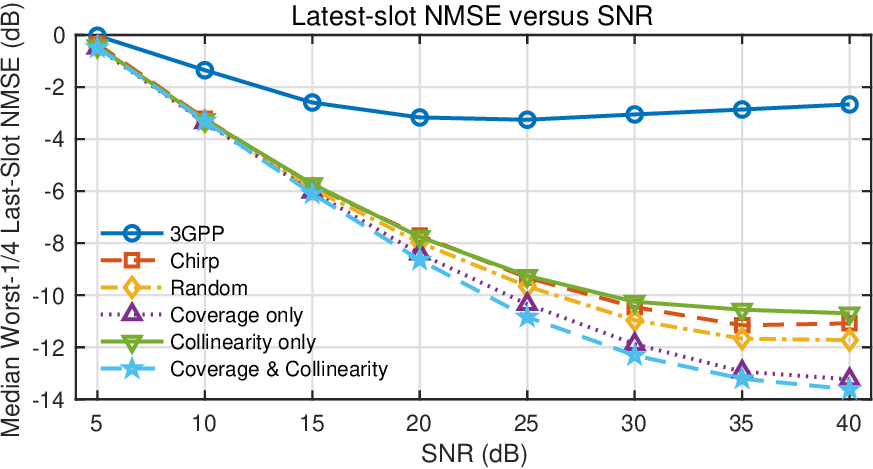}

\vspace{0.35em}

\includegraphics[width=\columnwidth]{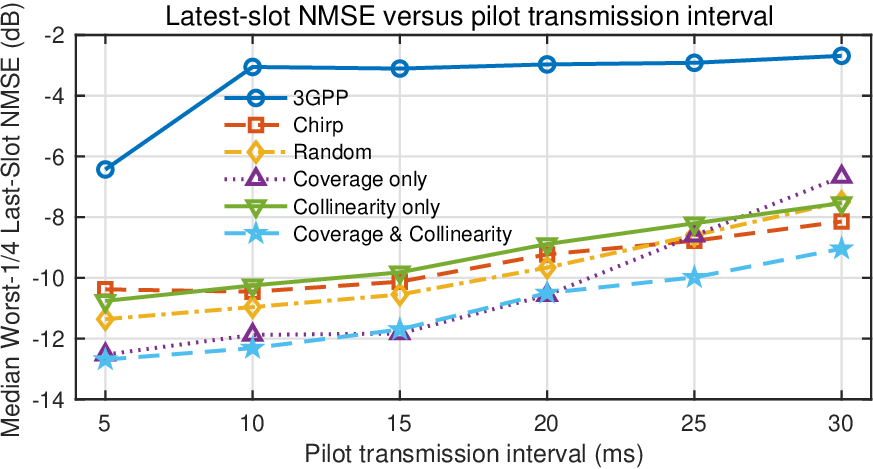}

\vspace{0.35em}

\includegraphics[width=\columnwidth]{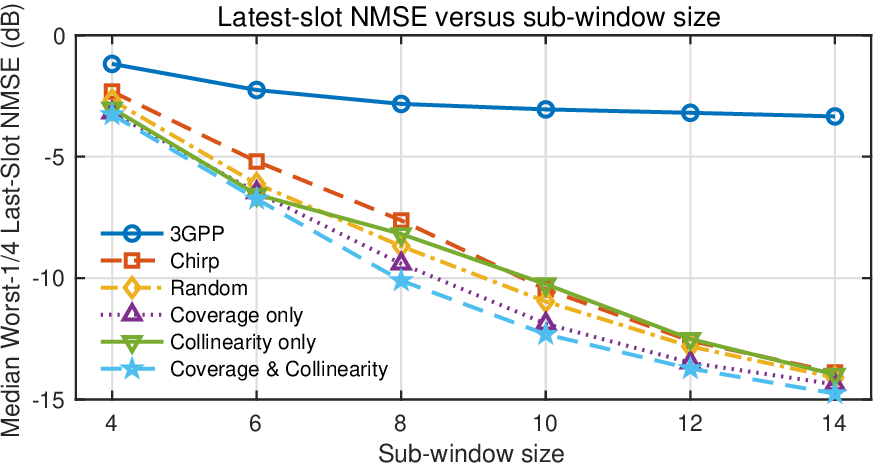}

\caption{Latest-slot recovery performance under sliding-window joint estimation for $k=17\ (M=24)$.
(a) NMSE versus SNR.
(b) NMSE versus pilot transmission interval.
(c) NMSE versus sub-window size.
The ``3GPP pilot'' refers to the pattern in Fig.~\ref{fig:pilot_patterns}.
The ``coverage only'' design removes \eqref{eq:col_penalty} and \eqref{eq:col_symmetry}, while the ``collinearity only'' design minimizes only the redundant-collinearity term subject to \eqref{eq:col_symmetry}.
``Coverage \& Collinearity'' denotes the proposed MCC pilot.}
\label{fig:numerical_results}
\end{figure}

Fig.~\ref{fig:numerical_results} shows that coverage is the primary design factor for latest-slot recovery: the coverage-only pattern remains competitive in the SNR and sub-window sweeps, whereas the collinearity-only pattern is consistently inferior. The benefit of collinearity control becomes clearer in harder regimes, especially as the pilot transmission interval increases. The sweep over $k$ further shows that some values, notably the composite case $k=16$, may induce unfavorable regular structure and visible performance loss. By comparison, the chirp pattern is motivated mainly by full-window coherence and is therefore less robust over short sub-windows. Overall, the proposed MCC pilot delivers the most stable performance across the tested settings.

\begin{figure}[t]
\centering
\includegraphics[width=\columnwidth]{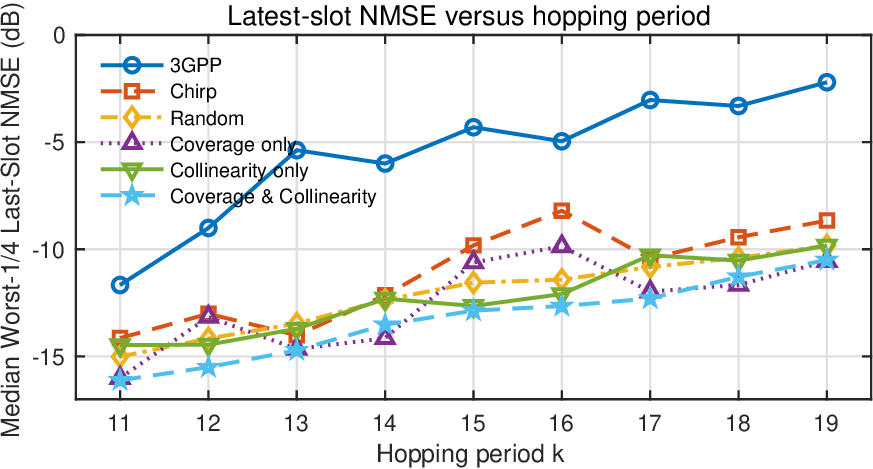}
\caption{Latest-slot NMSE versus hopping period $k\ (M=\lfloor 408/k\rfloor)$, with sub-window size $10$, SNR $=30$ dB, and pilot transmission interval $=10$ ms. The sweep includes both prime and composite values of $k$ to illustrate the robustness of different pilot patterns to the underlying modular structure.}
\label{fig:nmse_k}
\end{figure}

\section{Conclusion}\label{sec:conclusion}

This letter studied pilot-pattern design for sliding-window latest-slot recovery in TDD systems under practical contiguous-subband and fairness constraints. By shifting the design viewpoint from conventional slot-wise estimation to multi-slot joint recovery on the time--frequency grid, we showed that effective pilot patterns should be designed jointly across the observation window rather than slot by slot. This viewpoint leads to two main principles: coverage, which reflects the geometry of channel-extrapolation capability, and redundant-collinearity suppression, which mitigates unfavorable regular structure in more challenging regimes. Although these principles are realized through geometric surrogates on time--frequency point pairs that target worst-case pilot configurations, they yield designs that are computationally tractable and robust across different window sizes, pilot transmission intervals, and hopping periods.

\section*{Acknowledgment}

The authors acknowledge the support from National Key R\&D Program of China under grant 2021YFA1003301, and National Science Foundation of China under grant 12288101.

\bibliographystyle{IEEEtran}
\bibliography{IEEEabrv,mybibfile}

\end{document}